\documentclass[12pt]{article}

\usepackage{cite}
\usepackage{amsmath,amssymb,amsfonts}
\usepackage{algorithmic}
\usepackage{textcomp}
\usepackage{xcolor}
\usepackage{wrapfig}
\usepackage{graphicx}
\usepackage{booktabs,multirow}
\usepackage{adjustbox}
\usepackage{tikz}
\usepackage{mathtools}
\usetikzlibrary{quantikz}
\usetikzlibrary{external}
\usepackage{tikz}
\usepackage{dsfont}
\usepackage[margin=1in]{geometry}
\newcommand{\colvec}[2][.8]{%
  \scalebox{#1}{%
    \renewcommand{\arraystretch}{.8}%
    $\begin{bmatrix}#2\end{bmatrix}$%
  }
}
\usepackage[hyphens]{url}
\usepackage[colorlinks=true]{hyperref}

\def\BibTeX{{\rm B\kern-.05em{\sc i\kern-.025em b}\kern-.08em
    T\kern-.1667em\lower.7ex\hbox{E}\kern-.125emX}}
\title{Quantum Netlist Compiler (QNC)} 
\author{Shamminuj Aktar$^{1}$, Abdel-Hameed A. Badawy$^1$, Nandakishore Santhi$^2$}
\date{
    \small
    $^1$ Klipsch School of Electrical and Computer Engineering, New Mexico State University\\
    $^2$ Information Sciences Group, Los Alamos National Laboratory
}

\begin{document}
\maketitle
\thispagestyle{plain}
\pagestyle{plain}

\begin{abstract}
Over the last decade, Quantum Computing hardware has rapidly developed and become a very intriguing, promising, and active research field among scientists worldwide. To achieve the desired quantum functionalities, quantum algorithms require translation from a high-level description to a machine-specific physical operation sequence. In contrast to classical compilers, state-of-the-art quantum compilers are in their infancy. We believe there is a research need for a quantum compiler that can deal with generic unitary operators and generate basic unitary operations according to quantum machines’ diverse underlying technologies and characteristics. In this work, we introduce the Quantum Netlist Compiler (QNC) that converts arbitrary unitary operators or desired initial states of quantum algorithms to OpenQASM-2.0 circuits enabling them to run on actual quantum hardware. Extensive simulations were run on the IBM quantum systems. The results show that QNC is well suited for quantum circuit optimization and produces circuits with competitive success rates in practice.
\end{abstract}

\section{Introduction}
Quantum computing promises to exploit quantum mechanical properties to efficiently solve real-world problems that could not be solved using classical computers in polynomial time. Over the years, quantum algorithms have been getting overwhelming attention from scientific research communities due to their inherent theoretical performance promise over conventional computing paradigms~\cite{grover1996fast,deutsch1992rapid,biamonte2017quantum,han2002quantum,mavroeidis2018impact,li2019variational}. Tech giants like Google, IBM, Intel, and Microsoft are investing in the race of quantum processors. IBM, the first company to provide publicly-accessible superconducting qubit quantum machines, recently announced their 127-qubit quantum Eagle processor, considered a milestone in quantum technology~\cite{ibm-127}. The fundamental objective of running quantum algorithms on quantum computers is to solve various computational and scientific tasks more efficiently and faster than existing classical solutions.
However, there are several issues to consider when implementing quantum algorithms on real quantum computers. As the physical gate set of quantum systems differs from logical gates used in quantum algorithms, physical machines require the translation of logical gates into an actual physical gate set. The system’s connectivity is another essential issue as many quantum algorithms are typically designed for fully-connected hardware. However, quantum computers currently available have certain constraints on qubit connectivity~\cite{linke2017experimental}. As current quantum machines do not have enough qubits, many quantum compilers utilize several optimization techniques to reduce overall gate count and circuit depth~\cite{liu2021relaxed}. On the other hand, quantum programs need to be executed within the machine’s coherence time~\cite{divincenzo2000physical}. 

Quantum-computing researchers have made tremendous efforts toward identifying technical challenges and offering solutions to overcome those barriers. The Scaffcc compiler~\cite{javadiabhari2014scaffcc} is based on the LLVM framework, which allows compilation for large-scale applications and considers routing for specific hardware architecture. IBM Qiskit~\cite{ibm-qiskit} quantum software framework compiles user programs into QASM circuits targeting the IBM Q Experience. PyQuil~\cite{computing2019pyquil} is developed for Rigetti systems, and  Cirq~\cite{hancockcirq} is targeted for Google backends, while ProjectQ~\cite{steiger2018projectq} supports multiple hardware technologies. TriQ~\cite{murali2019full} takes gate sequence, qubit connectivity, and calibration data for the target machine as input and generates mappings specific to a machine's qubit topology. The full-stack quantum compiler (staq~\cite{amy2020staq}) takes QASM files with oracles, ancillae, and gate declarations and applies gate simplification, rotation folding, CNOT resynthesis, and qubit routing for circuit optimization. OpenQL~\cite{khammassi2021openql} quantum programming framework applies gate decomposition along with optimization techniques such as gate dependency analysis, sequence optimization, etc. Recently, Cambridge Quantum Computing Ltd.~proposed $t\ket {ket} $~\cite{sivarajah2020t} quantum software development platform, which supports circuit optimization, mapping to physical qubits, and noise-aware graph placement for qubit routing.

\begin{figure*}[tbp]
\vspace*{-3ex}
    \centering
    \includegraphics[width=0.82\linewidth] {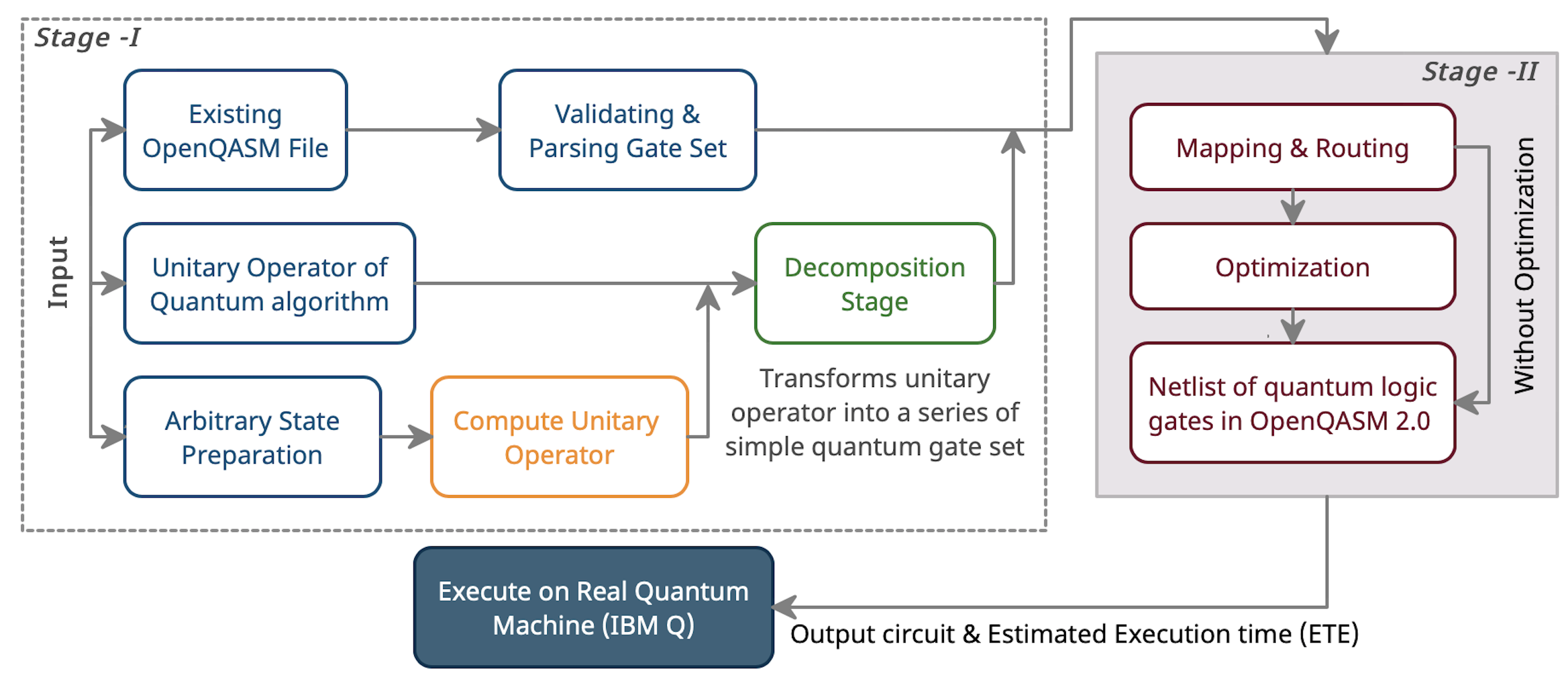}
\caption{
{\textbf{Overview of Quantum Netlist Compiler (QNC)} \label{Fig1}}}
\vspace*{-3ex}
\end{figure*}

In this paper, we introduce the Quantum Netlist Compiler (QNC), which translates unitary operators of a quantum algorithm into logical gate sequences or netlists according to a given machine topology. It also implements specialized routines for generating arbitrary states for quantum algorithms. We discuss the compiler's architecture and potential future research use in depth. We evaluate the accuracy of QNC by running the output circuits on different quantum machines. Experimental results show that the auto-generated circuits from QNC are competitive compared to many available compilers. QNC is a purpose-built research compiler written from scratch in C++-17 with a clean, lean code base, allowing easy future customization. We believe QNC is a research tool that will help researchers implement and debug their specific routines and new functionalities of quantum compilation and circuit optimization within QNC more efficiently.

The rest of this paper is organized as follows. Section~\ref{motivation} includes motivation of the work. Section~\ref{qnc} includes a detailed description of the Quantum Netlist Compiler (QNC), optimization techniques, and unique features. In Section~\ref{eval}, we discuss the experimental setup and results of QNC performance compared to conventional quantum compilers. Finally, Section~\ref{con} concludes the paper with possible future directions.

\section{Motivation}
\label{motivation}
For implementing a quantum algorithm in real quantum hardware, the first step is to find and build the quantum oracle which will be applied to the input state. A quantum oracle is essentially a unitary operator acting on $n$ qubits, which can be expressed as a $2^n \times 2^n$ unitary matrix. This unitary transformation needs to be constructed using the available primitive gates of particular quantum hardware. For this decomposition, at first, we need to apply a unitary diagonalization of the $2^{(n+1)} \times 2^{(n+1)}$ matrix using multi-qubit-controlled single-qubit unitary Given’s rotation operations \cite{adedoyin2018quantum}. Using standard techniques, it is possible to decompose such multi-qubit-controlled single-qubit operations further into primitive gate sets~\cite{nielsen2002quantum}. After this step, we get single and multi-qubit gates assuming full connectivity of the quantum hardware. Current IBM machines are not fully connected; they have linear nearest neighbor (LNN) connectivity. Thus, we will be left with arbitrary CNOT gates which do not respect the real topology of the target quantum machine. For LNN architectures, we need to further decompose the CNOT gates into a combination of available CNOT gates according to proper qubit connections. For example, suppose a particular quantum algorithm requires a CNOT gate between the $j^{th}$ \&  $l^{th}$ qubits. However, there is no connection between  the $j^{th}$ \&  $l^{th}$ qubits in the actual machine topology although qubits $j$ \& $k$ and $k$ \& $l$ are connected. Therefore, a CNOT gate between the $j^{th}$ \&  $l^{th}$ qubits can be constructed as follows:

\begin{equation}
    CNOT_{jl} = CNOT_{kl} CNOT_{jk} CNOT_{kl} CNOT_{jk} 
\end{equation}

Thus, manually decomposing the unitary operator of a quantum algorithm into an available gate set becomes very difficult and requires in-depth domain knowledge. Traditionally, we would need to decompose the unitary matrix by hand and compute the corresponding quantum circuit. Most state-of-the-art quantum compilers take quantum circuits as input, then apply routing, optimization, and other mappings to generate circuits for physical gate sets. The Quantum Netlist Compiler (QNC) starts with a unitary operator as input. Then, it automatically decomposes the unitary operator and transforms the operator into a series of simple quantum gate sets. Additionally, QNC allows preparing arbitrary states, which can be the input of many quantum algorithms. QNC computes the mathematical unitary operator from an arbitrary input state and then applies generic decomposition to generate the physical gate set. This extra ability to deal with the unitary operator as input sets QNC apart from most traditional compilers and fundamentally affords the quantum programmer more flexibility. It also makes QNC an interesting candidate compiler among other mainstream quantum compilers offering to decompose unitary operators. 

We show that QNC can generate competitive circuits for quantum algorithms and quantum state preparation. QNC implementation is simple and easily understandable
compared to traditional quantum compilers with multiple functionalities. We believe the simplicity of QNC will allow researchers to customize QNC efficiently with new features and routines.

\section{Quantum Netlist Compiler (QNC)}
\label{qnc}
Quantum Netlist Compiler (QNC) is designed to be an extensible compiler infrastructure capable of processing a quantum algorithm in its most basic form: a unitary linear operator, an arbitrary starting state, or even an existing OpenQASM circuit for further optimizations. QNC transforms operators into a sequence of simple quantum gates, using OpenQASM as a quantum intermediate representation (QIR) -- resulting circuits can then be run on existing quantum computers. QNC achieves this objective using mathematically robust decomposition, optimization, and graphical routing techniques, as described below. 
 
\subsection{Overall Architecture}
The operation of the QNC compiler can be decomposed into three distinct steps, which are organized within two stages as in Figure~\ref{Fig1}: (1) Decompose either an initial state vector or a unitary operator into simpler unitary operator sequences which are directly implementable as qubit gates on physical quantum computers, (2) Map all logical qubits to physical qubits, and graphically route interacting qubits so that they are physical neighbors in the target quantum computer topology (3) Coalesce and simplify adjacent quantum gates iteratively to reduce overall circuit complexity. Suppose an existing quantum circuit is already known for an algorithm; then, QNC can optionally avoid doing step (1), thereby only performing map, route, and optimize a quantum circuit in a scalable fashion.

\subsection{Input to QNC}  
\subsubsection{Unitary operator from Quantum Algorithm}

In quantum physics, a unitary operator represents transformation of a quantum state, \textit{e.g.}, a quantum initial state $\ket{\phi}$ becomes state $\ket{\psi}$ after $\ket{\psi} = U \ket{\phi} $, where $U$ is the unitary transformation. Any quantum evolution can be represented as unitary operator, which is the most basic and generic form. Given a unitary operator from a quantum algorithm as an input, QNC decomposes the matrix into a physical gate-set according to the machine topology. The input unitary matrix must be in the form of space-separated list of complex or real numbers. The complex numbers are represented as a tuple within brackets, \textit{e.g.}, $(1.0, 2)$ stands for complex value $1.0 + 2.0i$. QNC checks if the unitary operator is compatible with the provided logical to physical qubit mapping. It also computes the number of qubits required for generating an equivalent circuit. An example of the unitary transformation matrix for two qubit Quantum Fourier Transformation (QFT) is \\
    \begin{center}
        $
        F_2 =\dfrac{1}{2}
        \begin{bmatrix}
        1 &  1 & 1 & 1 \\
        1 &  i & -1 & -i \\
        1 &  -1 & 1 & -1 \\
        1 &  -i & -1 & i 
        \end{bmatrix} 
        $ \\
    \end{center}
    
    QNC takes this unitary in the following format: \\
    \begin{center}
        $
        F_2 =
        \begin{bmatrix}
        0.5 &  0.5 & 0.5 & 0.5 \\
        0.5 & (0.0,0.5) & -0.5 & (0.0,-0.5) \\
        0.5 & -0.5 & 0.5 & -0.5 \\
        0.5 & (0.0,-0.5) & -0.5 & (0.0,0.5) \\
        \end{bmatrix} 
        $
    \end{center}        

\subsubsection{Initial State Preparation}
State vectors describe a particular state of a quantum system. Many quantum algorithms require the system to be initialized to an arbitrary state (\textit{e.g.}, GHZ state~\cite{cruz2019efficient}, Dicke state~\cite{aktar2022divide}, etc.  instead of a ground state. To generate an $n$-qubit state, we need to calculate the unitary transformation, which converts the $ N $-dimensional vector to the desired state. Thus, if we start from $\ket{0}$ as the initial qubit state, which is the default according to the physical machine’s architecture, the desired state can be calculated as follows:

\begin{equation}
 {U_s} \ket{0} \rightarrow \ket{s}
\end{equation}

where $U_s$ is the unitary operator for the state preparation, $\ket{0}$ is the initial qubit state, and $\ket{s}$ is the desired initial state. Given any $n$-qubit desired state-vector as input, QNC automatically calculates the required unitary transformation and generates a circuit to prepare the specified state-vector as output. For example, Greenberger–Horne–Zeilinger (GHZ)~\cite{cruz2019efficient} state is a maximally entangled quantum state of at least three qubits. It is a superposition of all $n$ qubit states with all subsystems being in zero states or all subsystems in one states. Three qubits GHZ state is,

\begin{align}
    \ket{GHZ} = \frac{\ket{000}+\ket{111}}{\sqrt{2}}
\end{align}

An equivalent quantum circuit to prepare GHZ state can be generated using QNC from the input initial state vector.

\subsubsection{Permutation Operator}
The permutation operator permutes two or more subsystems. QNC Reads a permutation operator from the given file in Cauchy's single-line notation. It will decompose the permutation operator into a sequence of swap operations.

\subsubsection{QASM Input For Further Optimization}

QNC offers to input well-known quantum circuits directly instead of starting from a unitary operator or initial state. Given quantum circuits in the form of an OpenQasm file, QNC will apply optimization techniques and perform qubit mapping and routing according to the user-specified machine topology. This feature will be helpful in quickly generating optimized and correctly mapped circuits for any quantum circuit.

 \subsection{Unitary Decomposition}
        
 A unitary linear operator can always be represented as a complex matrix. If the operator happens to be a permutation operation, QNC handles it slightly more efficiently, utilizing its specific structure. Otherwise, the qubits are ordered naturally or encoded using a Gray code to minimize control-NOT gate sequences during pivoting in the following Givens decomposition operation. The unitary matrix is decomposed into simple controlled qubit gates using a recursive Givens decomposition operation. QNC proceeds slightly differently if used to perform an initial state preparation. Given a desired initial qubit state $\ket{s}$ other than $\ket{0}$, QNC will first compute a unitary operator to convert $\ket{0}$ to $\ket{s}$.  The rest of the decomposition operations are performed as described below. However, if the input to QNC is already an OpenQASM circuit, then this step is skipped.
        
\subsubsection{Qubit Encoding}
To generate the two-level unitaries, we choose basis vectors $\ket{e_k} = \bigotimes_{i} \ket{x^k_i}$ where $i ={1,..n}$ and $x^k_i \in {\{0,1\}} $. As the order of the basis vectors is not fixed, QNC offers natural encoding and uses a Gray code to minimize control-NOT gate sequences during pivoting in the following Givens decomposition operation. Gray code is a palindrome-like ordering of binary numbers where adjacent elements $gr^n_i$ and $gr^n_{i+1}$ differ only in one bit. Table~\ref{tbl:I} illustrates the natural and gray coding of basis vectors for n =4.
    
\subsubsection{Givens Decomposition}
 A unitary matrix $U$ can be decomposed into a product of two-level unitary matrices using Givens rotation~\cite{li2013decomposition}. From unitary matrix $U$, we can find $G_i$ such that 

\begin{equation}
    (\Pi_{i=1..m} G_i).U = I \\ \longrightarrow  (\Pi_{i=m..1} G_i') = U
    \label{eq:givens}
\end{equation}

A Givens rotation~\cite{li2013decomposition} $G^{j,k}_i$ is a two-level matrix that represents a counterclockwise rotation on two basis vectors, $\ket{e_j}$ and  $\ket{e_k}$. Basically, the objective of the Givens rotation $G^{j,k}_i$ is to nullify $u_{j,i}$  using $u_{k,i}$ of matrix $U$. The following $G^{j,k}_i$ matrix when applied to an arbitrary unitary matrix $U$, will change $u_{j,i}$ to zero, if it is non-zero.

\begin{align}
    G^{j,k}_i = \begin{bmatrix}   
                        g^{k,k}_i & g^{k,j}_i \\   g^{j,k}_i & g^{j,j}_i \\
                \end{bmatrix} = \frac{1}{\sqrt{{|u_{j,i}|}^2+{|u_{k,i}|}^2}}\begin{bmatrix} 
                    u^{*}_{k,i} & u^{*}_{j,i} \\   -u_{j,i} & u_{k,i} \\
                \end{bmatrix}
\end{align}

In this way, we apply Givens rotation recursively to get the identity matrix on the left-hand side of equation~\ref{eq:givens} 
and get the product of Givens rotation matrices on the right. As $G^{j,k}_i$ are all two-level unitary matrices, their inverse are also two-level unitary matrices. For $n$ qubit initial state preparation, we perform the following transformation, where we calculate $\Pi_{i=1..m} G_i$ as discussed above,

\begin{equation}
    [(\Pi_{i=1..m} G_i).s = e_0] \mapsto [(\Pi_{i=m..1} G_i').e_0 = s]
\end{equation}
where $e_0  = [1,0..n]$. 
    
\begin{table}[t]
    \centering
    \caption{Natural and Gray coding of basis vectors for $n$=4.}
    \begin{adjustbox}{max width=0.99\linewidth,center}
    \begin{tabular}{lccc}
    \toprule
        \textbf{i} & \textbf{Basis Vector} &\textbf{Natural Code} & \textbf{Gray Code}\\
        \textbf{} & {}& {$n_r = i$} & \ ${g_r = iXOR(i/2)}$ \\
            \midrule
        0 & 00 & 00 & 00\\
        1 & 01 & 01 & 01\\
        2 & 10 & 10 & 11\\
        3 & 11 & 11 & 10\\
    \bottomrule
    \end{tabular}
    \end{adjustbox}
    \label{tbl:I}
\end{table}    
    
\subsubsection{Example}
In this section, we demonstrate the decomposition of a $4 \times 4$ unitary matrix of a swap gate~\cite{quantum-logic}. 
For simplicity, we are showing only the real values of the matrix. The unitary matrix for a SWAP gate is:

\begin{align}
   U = \begin{bmatrix}  
                1 & 0 & 0 & 0\\
                0 & 0 & 1 & 0\\
                0 & 1 & 0 & 0\\
                0 & 0 & 0 & 1
                \end{bmatrix}
\end{align}

Applying Gray coding to the basis vector set, the new row encoding becomes $g_r = [0,1,3,2]$. We will apply Givens rotation to $U$ recursively until it becomes an identity matrix and generates two-level unitary matrices. Starting from $i = g_r[0] = 0$ (first column), we find that no changes required for first column, \textit{i.e.}, elements $u_{1,0}, u_{2,0}, u_{3,0}$ are already zeros. Then, we move forward with $i = g_r[1] = 1$ (second column) and find that our first objective is to null $u_{2,1}$. It can be done by applying the following rotation using $u_{3,1}$ 

\begin{align}
    G^{2,3}_1 = \frac{1}{\sqrt{{|u_{2,1}|}^2+{|u_{3,1}|}^2}}
                \begin{bmatrix} 
                    u^{*}_{3,1} & u^{*}_{2,1} \\
                    -u_{2,1} & u_{3,1} \\ 
                \end{bmatrix}
\end{align}

 Putting the values from $U$ we get,  
$
G^{2,3}_1 = \begin{bmatrix} 
                    0 & 1 \\   
                    -1 & 0 \\ 
            \end{bmatrix} $ \\
and ${G_1}U = U_{r} =
            \colvec{
                1 & 0 & 0 & 0\\
                0 & 0 & 1 & 0\\
                0 & 0 & 0 & -1\\
                0 & 1 & 0 & 0
            } 
$, where $u_{2, 1}$ is now zero. \\ \\
Next, the objective is to null the $u_{3, 1}$ element in the reconstructed matrix $U_r$ using $u_{1, 1}$ by applying the rotation $
    G^{3,1}_1 = \begin{bmatrix} 
                    0 & 1 \\   
                    -1 & 0 \\ 
                \end{bmatrix} $
and get ${G_2}{G_1}U = U_r = \colvec{ 
                1 & 0 & 0 & 0\\
                0 & 1 & 0 & 0\\
                0 & 0 & 0 & -1\\
                0 & 0 & -1 & 0
                }$ \\ \\
where, $u_{3, 1}$ is now zero.    
Similarly, we proceed for $i = g_r[2] = 3$ (last column), and look for possible Givens rotation for the reconstructed matrix $U_r$. We find the objective is to null $u_{2, 3}$ using $u_{3, 3}$ by applying the rotation 
$
G^{2,3}_3 = \begin{bmatrix} 
                    0 & -1 \\   
                    1 & 0 \\ 
                \end{bmatrix} $
and ${G_3}{G_2}{G_1}U = U_r = 
                \colvec{ 
                1 & 0 & 0 & 0\\
                0 & 1 & 0 & 0\\
                0 & 0 & -1 & 0\\
                0 & 0 & 0 & 1
                } 
    $ \\
where $u_{2, 3}$ is now zero. Finally we check for $i = g_r[3] = 2$ (third column) and find that we already have zeros in $u_{0,2}, u_{1,2}$ and $u_{3,2}$ of $U_r$. Finally, we check if the product of the diagonal elements of the generated matrix is 1. At this step, only the $gr[(N-1)]$ diagonal element can be non-unity for unitary operators, and only the $gr[0]$ element can be non-unity for state preparation. 
    
Here, we find that the diagonal product is $-1$ as the reconstructed matrix $U_r$ contains negative value in third column, \textit{i.e.}, $i = gr[3] = 2$. Now, our goal is to use the $u_{0, 2}$ entry to make the $u_{2,2}$ entry 1 by applying the rotation
    $
    G^{0,2}_2 = \begin{bmatrix} 
                        -1 & 0 \\   
                        0 & 1 \\ 
                    \end{bmatrix} $
    and ${G_4}{G_3}{G_2}{G_1}U = 
                    \colvec{ 
                    1 & 0 & 0 & 0\\
                    0 & 1 & 0 & 0\\
                    0 & 0 & 1 & 0\\
                    0 & 0 & 0 & 1
                    } = \mathds{1} $
    
Finally, we get the identity matrix on the left side and from the definition of a unitary matrix, our decomposition becomes $U = {G_1}{G_2}{G_3}{G_4}$, \textit{i.e.},

  \begin{align}
      U =
                  \colvec{ 1 & 0 & 0 & 0 \\ 
                    0 & 1 & 0 & 0 \\
                    0 & 0 & \boldsymbol{0} & \boldsymbol{-1} \\ 
                    0 & 0 & \boldsymbol{1} & \boldsymbol{0}}
                 \colvec{
                    1 & 0 & 0 & 0 \\ 
                    0 & \boldsymbol{0} & 0 & \boldsymbol{1} \\
                    0 & 0 & 1 & 0 \\ 
                    0 & \boldsymbol{-1} & 0 & \boldsymbol{0}}
                \colvec{
                    1 & 0 & 0 & 0 \\ 
                    0 & 1 & 0 & 0 \\
                    0 & 0 & \boldsymbol{0} & \boldsymbol{1} \\
                    0 & 0 & \boldsymbol{-1} & \boldsymbol{0}}
              \colvec{
                    \boldsymbol{1} & 0 & \boldsymbol{0} & 0 \\ 
                    0 & 1 & 0 & 0 \\
                    \boldsymbol{0} & 0 & \boldsymbol{-1} & 0 \\ 
                    0 & 0 & 0 & 1}
\end{align}


\begin{figure*}[tbp]
\vspace*{-3ex}
\centering
\includegraphics[width=\linewidth] {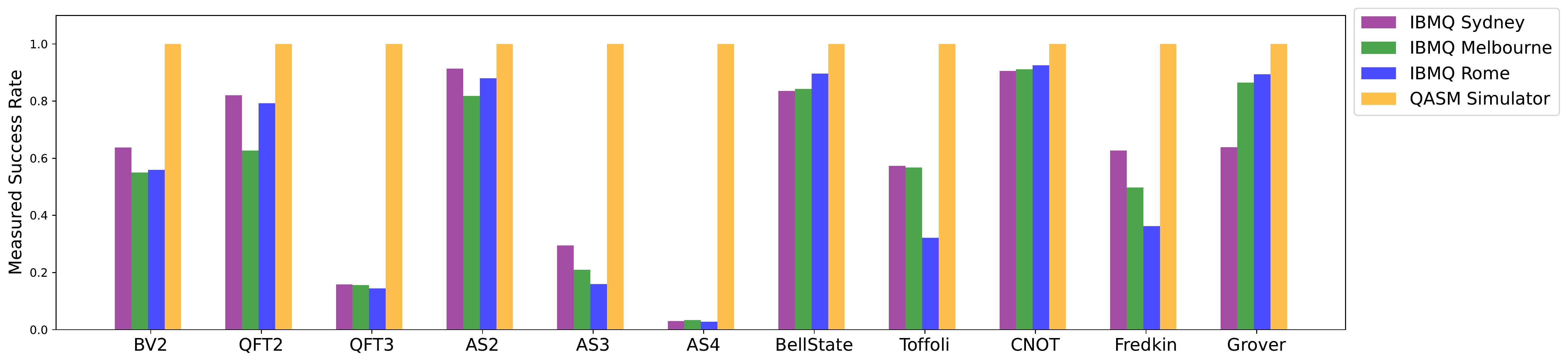}
\caption{The success rate of all benchmarks on IBM machines (IBMQ27 Sydney, IBMQ15 Melbourne, and IBMQ5 Rome) quantum processors and IBM QASM Simulator. The 100\% success rate for all benchmarks on the QASM simulator is indicative of the accuracy of the generated QNC circuits. For BV4, BV6, and QFT4, we only show the simulator’s results and skip execution on actual quantum machines as the circuits are too big and the estimated execution time (ETE) exceeds the coherence times of the quantum machines.}
\label{fig-all}
\end{figure*}

\subsubsection{Mapping and Qubit Routing}
Present quantum devices have certain constraints on qubit connectivity as the qubits do not have full connectivity. These constraints are not considered in the logical quantum circuits and must be modified before executing on the machines. The physical qubits of the quantum machines can interact only with their neighbors as per the machines' topology. We define the qubits used in a quantum circuit as logical qubits and the actual machine's qubits as physical qubits. A quantum machine connectivity can be represented using an undirected graph $G = ({Q},{E})$ where vertices ${Q}$ denote physical qubits, edges ${E}$ denote connections between qubits. For multi-qubit operations, it is essential to make sure the logical qubits of the circuits are mapped to physical qubits. 


    
\begin{figure}[t!]
   
    \centering
    \resizebox{0.5\columnwidth}{!}{%
        \begin{tikzpicture}[
node distance=1.5cm, scale=4,  align=center, 
state/.style={circle, draw=blue, minimum size=3em}]
        \node[state,font=\fontsize{15pt}{15pt} ] (0)  { $q_i$};
        \node[state,font=\fontsize{15pt}{15pt} ] (1) [right of=0]  {$q_j$}; 
        \node[state,font=\fontsize{15pt}{15pt} ] (2) [right of=1] {$q_k$};
        \node[state,font=\fontsize{15pt}{15pt} ] (3) [right of=2] {$q_l$}; 
        \node[state,font=\fontsize{15pt}{15pt} ] (4) [right of=3] {$q_m$};
        
        \draw (0) -- (1);
        \draw (1) -- (2);
        \draw (2) -- (3);
        \draw (3) -- (4);
         \draw[dashed,bend left=15]  (1) to[out=-90,in=-70] (3) node [fill=none, below right ] {CNOT};
        
        \end{tikzpicture} 
    }
    \newline
    \resizebox{0.99\columnwidth}{!}{%

        \begin{quantikz}
            \ket{q_j}&  \ctrl{2} & \midstick[3,brackets=none]{=}\qw& \ctrl{1} & \qw & \ctrl{1} & \qw  & \midstick[3,brackets=none]{=}\qw& \qw& \ctrl{1}& \qw& \ctrl{1}& \qw\\
            \ket{q_k}&  \qw      & \qw                             & \targ{1} & \ctrl{1} & \targ{1} & \ctrl{1}& \qw & \ctrl{1}&\targ{}& \ctrl{1}&\targ{}& \qw\\
            \ket{q_l}&  \targ{}  & \qw                             & \qw & \targ{1} & \qw & \targ{1}& \qw&\targ{}&\qw&\targ{}&\qw& \qw
        \end{quantikz}  
    }
    \caption{Implementation of the CNOT gate between qubit $q_j$ \& $q_l$ when $q_j$ \& $q_l$ are not connected but there is connectivity between $q_j$ \& $q_k$ and  $q_k$ \& $q_l$. }
    \label{cnot-route}
\end{figure}
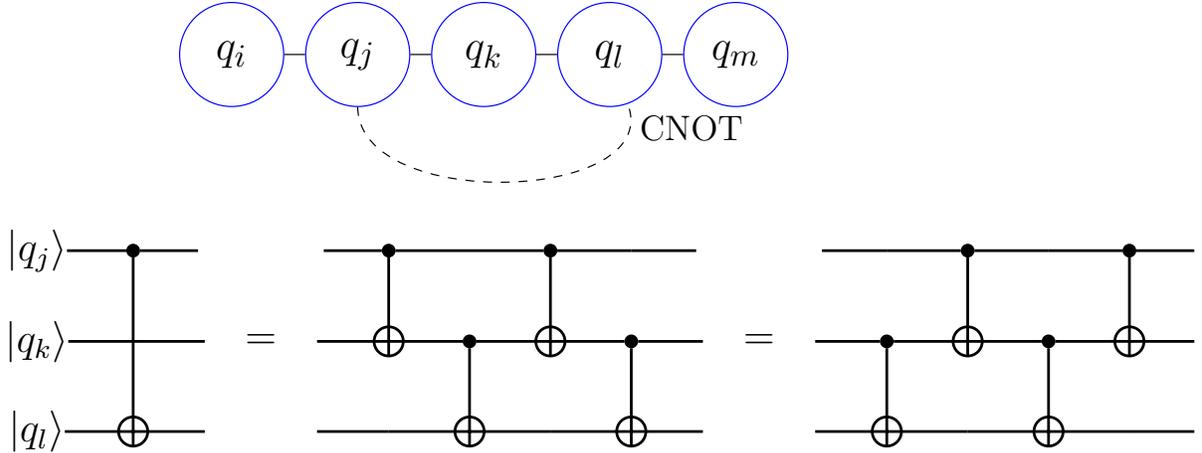

Figure~\ref{cnot-route} shows an example implementation of the CNOT gate among qubits that are not immediately connected. In QNC, the logical qubits are mapped to physical qubits using a user-provided permutation map of the machine. At first, it pre-computes optimal paths for qubit routing based on the given topology. QNC applies Floyd–Warshall algorithm~\cite{weisstein2008floyd} to compute the shortest paths between each pair of vertices ${Q}$. Specifically, the objective is to find the shortest path between ${q}_i$ and ${q}_l$ using any vertex in ${q}_N$ where $N = {1,2,...,n}$. Therefore, QNC will try to route individual qubits using the shortest paths, and once the operation is performed, QNC will route the qubit back to its original position. 

\subsubsection{Optimization Step} In this step currently, QNC coalesces adjacent quantum gates whenever feasible to reduce the overall gate count and hence the execution time. Additional optimizations are planned for the future.

\subsection{Compiler Output and Timing Analysis} After performing the optimizations, QNC writes out the final quantum circuit as a netlist of quantum logic gates in the OpenQASM-2.0 language. QNC will also model the execution time of each quantum gate on IBM machines using the estimated time periods required for Frame-Changes (FC), Gaussian-Derivatives (GD), and Gaussian-Flattops (GF). The user can compare the estimated execution time with the quantum computer design coherence time and thus evaluate the quantum circuit success probability produced by the compiler.






\begin{figure*}[tbp]
\vspace*{-3ex}
    \centering
    \includegraphics[width=0.9\linewidth] {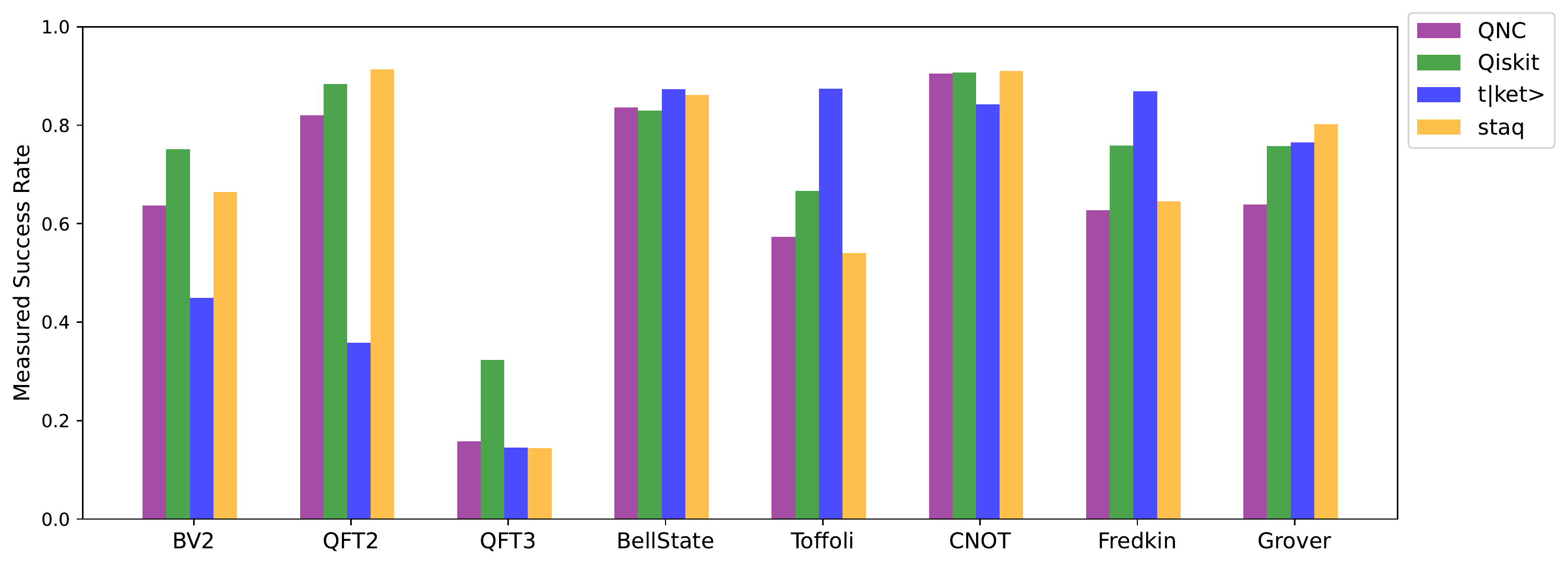}
\caption{
Evaluation of QNC performance against other conventional compilers in terms of measured success rate and the circuits runtime. In (a), we demonstrate the comparative success rate of benchmarks using QNC, IBM Qiskit, $t\ket {} $, and the staq compilers on 27 qubit IBMQ Sydney quantum processor. The comparison shows that QNC generates a competitive success rate for all benchmarks compared to conventional compilers.}
\label{fig2}
\vspace*{-3ex}
\end{figure*}

\section{Performance Evaluation}
\label{eval}
Manually  decomposing  a  unitary  operator  of  a quantum  algorithm into an  available gate set is laborious and requires in-depth domain knowledge. QNC can take a unitary matrix operator as input for a specific quantum problem and outputs the corresponding optimized quantum gate circuits depending on the target physical device topology. Since the QNC compiler deals with generic unitary operators as input, it means that certain circuits will unavoidably need $\mathcal{O}(4^n)$ simple gates for an $n$ qubit operation~\cite{nielsen2002quantum}. A generic compiler like QNC cannot do much to reduce the gate count for a generic operator in the worst case. When quantum circuits are hand-tuned or derived using symmetry considerations. For example, in specific cases, the gate count required can be brought down to the order of a polynomial in the number of qubits. 
This gap is expected to narrow or perhaps even reverse as special cases are handled appropriately in Stage-I of compilation in the future. 
While hand-tuned circuits can sometimes have far fewer gates in their implementation, it also usually comes at a high price since hand-tuning is very time-consuming and error-prone.

We perform simulations using QNC-generated quantum gate circuits for various problems and validate the correctness and competitiveness compared to leading tools available to the research community.
We use relatively small benchmarks, including the Bernstein-Vazirani~\cite{bernstein1997quantum} algorithm for two, four, and six qubits (BV2, BV4, BV6), state preparation for two, three, and four qubits (AS2, AS3, AS4), Quantum Fourier transformation~\cite{qft} for two, three and four qubits (QFT2, QFT3, QFT4), two-qubit Grover algorithm~\cite{grover}, the Bell state, and multi-qubit quantum gates such as CNOT, Toffoli and Fredkin gates~\cite{multi-gate}. These benchmark problems represent essential parts of many quantum algorithms.

The QNC compiler was run on a 1.4 GHz quad-core Intel Core i5 processor with 8GB of RAM running macOS Big Sur (v11.6.2) to generate the quantum circuits.

Figure~\ref{fig-all} shows the success rate of the benchmarks on various quantum processors (IBMQ27 Sydney, IBMQ15 Melbourne, \& IBMQ5 Rome) and IBM QASM Simulator. We can see that the simulator always generates a 100\% probability of finding the desired state for all benchmarks. This reassures the accuracy of the QNC-generated circuits. We did not run BV4, BV6, and QFT4 on real IBM machines since their estimated execution time would have exceeded the IBM quantum machine coherence time(s). This is because QNC starts with a unitary matrix without any prior knowledge about a specific quantum circuit. Thus, the generic decomposition of the unitary operator may generate a large number of gates. We can see in Figure~\ref{fig-all} that for QFT3, AS3, and AS4, the actual success rates are much lower than the expected probability. This is because these circuits (3-qubit quantum Fourier transformation and 3-qubit, \& 4-qubit arbitrary state preparation) have very long estimated execution times. Furthermore, the success rate of the IBMQ5 Rome for almost all the benchmarks (except QFT3, AS3, and AS4) is higher than other machines. 

\textbf{Observation I:} \textit{QNC decomposes the input unitary matrix and generates an optimized simple gate set based on the target machine topology. All the circuits generated from QNC are 100\% accurate on the QASM simulator. Sometimes QNC may generate a longer circuit which can generate noisy output states because of the coherence time limitation of the particular Quantum processor.}

\subsection{Performance Comparison with Existing Compilers}
To assess the performance of QNC against conventional quantum compilers, we also generated circuits using IBM Qiskit~\cite{ibm-qiskit}, $t\ket{ket}$~\cite{sivarajah2020t}, and the staq~\cite{amy2020staq} compiler. We choose Qiskit, $t\ket{ket}$, and staq since they represent the latest, most popular, and publicly available toolchains supporting the OpenQASM standard, similar to QNC. We did not include other compilers since they were either non-configurable or targeted for a different quantum architecture. Currently, the staq compiler is hard-coded for a few IBMQ machines, which are retired now. As the interface of staq is not user-friendly, we added the topology of the latest IBMQ machines into staq to evaluate the performance of staq.  Figure~\ref{fig2} shows a comparative study of the measured success rate for all benchmarks on the IBMQ27 Sydney quantum machine using QNC, IBM Qiskit, $t\ket{ket}$, and staq. We can clearly notice that the QNC implementations exhibit   competitive performance for circuits such as BV2, QFT2, BellState, and Grover (2-qubit) like IBM Qiskit and $t\ket{ket}$ implementations. Since QNC produces circuits from unitary matrices, the circuit length is slightly larger for multi-qubit quantum gates, which justifies the Toffoli and Fredkin gates measured outcome. This comparison is intended to highlight how QNC can be a competitive compiler in terms of the measured success rate for the desired state. 

\textbf{Observation II:} \textit{Although QNC may generate a large number of simple gates, the success probabilities of the outputs for the QNC-generated circuits are almost close or competitive to the IBM Qiskit circuits with the highest optimization applied. 
}

\section{Conclusion \& Future Work}
\label{con}
In this paper, we introduced the QNC compiler along with its notable features and optimization techniques. We have open-sourced QNC for research purposes. Benchmarking on various quantum machines and comparative study against conventional quantum compilers showed competitive success rates of QNC-generated circuits. Several enhancements to QNC are planned in the near future, such as the ability to optimize for specific classes of unitary matrices so as to reduce the required number of gates. In addition, QNC can serve as a testbed for incorporating features like optimizations of external OpenQASM circuits by utilizing the optimization techniques of QNC, preparing quantum circuits for arbitrary states required for input to quantum algorithms, and using timing analysis from QNC outputs as a metric for optimizing quantum algorithms. In the hope that QNC will be a useful tool for quantum computation researchers, our implementation of QNC is available at \href{https://github.com/pujyam/QNC}{\underline{https://github.com/pujyam/QNC}}.

\section*{Acknowledgments} 
The authors thank the anonymous reviewers for their time and dedication. We thank IBM for giving us access to their IBM Q systems. This work is partially supported by Triad National Security, LLC subcontract \#581326. The views expressed in this paper do not necessarily represent the views of the DOE or the U.S. Government. Approved for public release LAUR\# LA-UR-22-28503.

\newpage
\bibliographystyle{plainurl}
\bibliography{main}
\end{document}